\RequirePackage{fix-cm}
\documentclass[smallextended, referee]{svjour3}       
\smartqed  
\usepackage[pdftex]{graphicx}
\usepackage{mathptmx}      
\usepackage{latexsym}
\usepackage{url}
\usepackage{cite}
\usepackage{array}
\usepackage{subfig}
\usepackage{amsmath}
\begin{document}

\title{Progress towards an electro-optical simulator for space based, long arms interferometers
}


\author{Pierre Gr\"uning         \and
        Hubert Halloin \and
        Pierre Prat  \and
        Sylvain Baron \and
        Julien Brossard \and
        Christelle Buy \and
        Antoine Petiteau 
}


\institute{P. Gr\"uning \at
             APC, AstroParticule et Cosmologie, Universit\'e Paris Diderot, \\
             CNRS/IN2P3, CEA/Irfu, Observatoire de Paris, Sorbonne Paris Cit\'e, \\ 
             10, rue Alice Domon et L\'eonie Duquet, \\
             75205 Paris Cedex 13, France
              Tel.: +33 1 57 27 60 77 \\
              Fax: +33 1 57 27 60 70\\
              \email{pierre.guning@apc.univ-paris7.fr}           
}

\date{Received: date / Accepted: date}

\maketitle

\begin{abstract}
We report the progress in the realization of an electronic / optical simulator for space based, long arm interferometry and its application to the eLISA mission. The goal of this work is to generate realistic optics and electronics signals, especially simulating realistic propagation delays. The first measurements and characterization of this experiment are also presented. With the present configuration a modest $10^6$ noise reduction factor has been achieved using the Time Delay Interferometry algorithm. However, the principle of the experiment has been validated and further work is ongoing to identify the noise sources and optimize the apparatus.

\end{abstract}

\section{Introduction}
\label{intro}
The evolved Laser Interferometer Space Antenna is a space-based project aiming at detecting gravitational waves in the frequency range 0.1 mHz to 10 Hz. eLISA consists of 3 spacecraft in a nearly-equilateral configuration, constantly following free-falling masses located at their center and orbiting around the sun\cite{ELISAWP2013}. The eLISA mission has recently been submitted as an L-class mission to the European Space Agency. The eLISA concept follows the LISA \cite{LISAYB2011,Danzmann2000} and NGO \cite{NGOYB2012} projects, for which extensive assessment studies have been made.

\begin{figure*}[h]
\centering
\subfloat[The eLISA constellation orbiting around the sun.]{\includegraphics[width=2.2in]{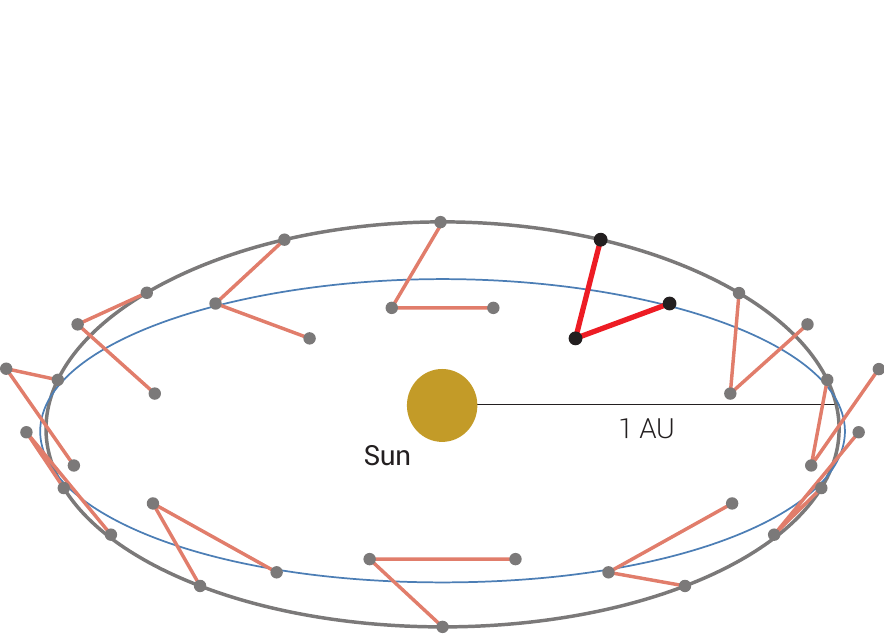}
\label{Fig:eLISAOrbits}}
\hfil
\subfloat[The V-shaped constellation of eLISA.]{\includegraphics[width=2.2in]{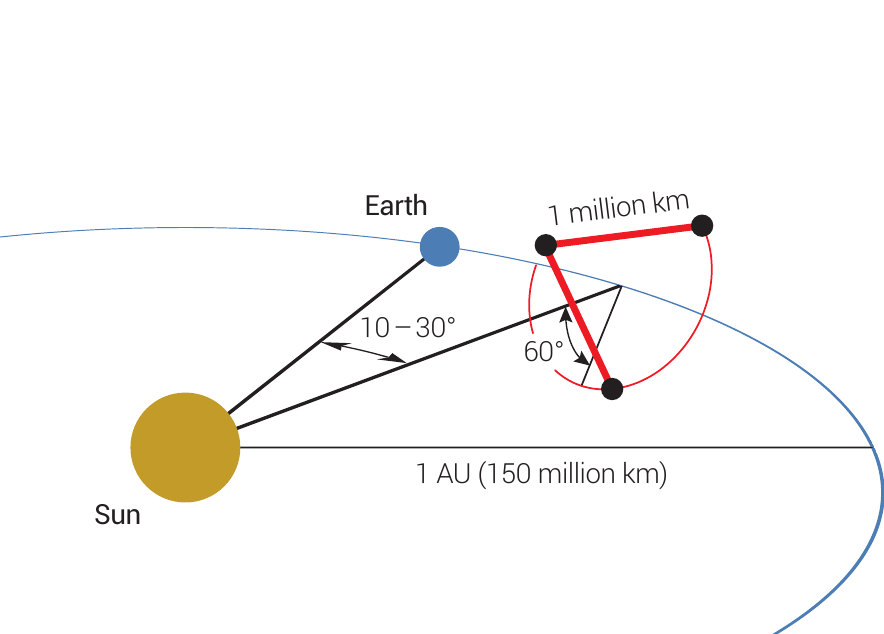}
\label{Fig:eLISAConstellation}}

\caption{Configuration of the eLISA orbits}
\label{Fig:eLISAMission}
\end{figure*}

In the eLISA concept, a 'mother' spacecraft is located at the vertex of a V-shaped configuration, with two 'daughter' spacecraft at the end of the two arms. Laser beams are propagating along the arms, effectively forming a Michelson-type interferometer with $10^{6}$~km arm lengths.  The spacecraft
follow independent heliocentric orbits without any station-keeping and form a nearly equilateral triangle in a plane that is inclined by 60¡ against the ecliptic (see figure~\ref{Fig:eLISAMission}).

These interferometric measurements are used to precisely monitor the distance between the inertial masses and, hence, to detect the tiny variation due to the pass of a gravitational wave.  The goal of LISA is to detect gravitational deformation as small as $\Delta L/L \approx 10^{-20}\sqrt{\text{Hz}}$ (i.e 10 pm per million of km) around 5 mHz.

This expected performance of eLISA relies on two main technical challenges: the ability for the spacecraft to precisely follow the free-flying masses and the outstanding precision of the phase shift measurement. 

\begin{figure*}[h]
\centering
\subfloat[Scheme of the eLISA interferometer.]{\includegraphics[width=2.2in]{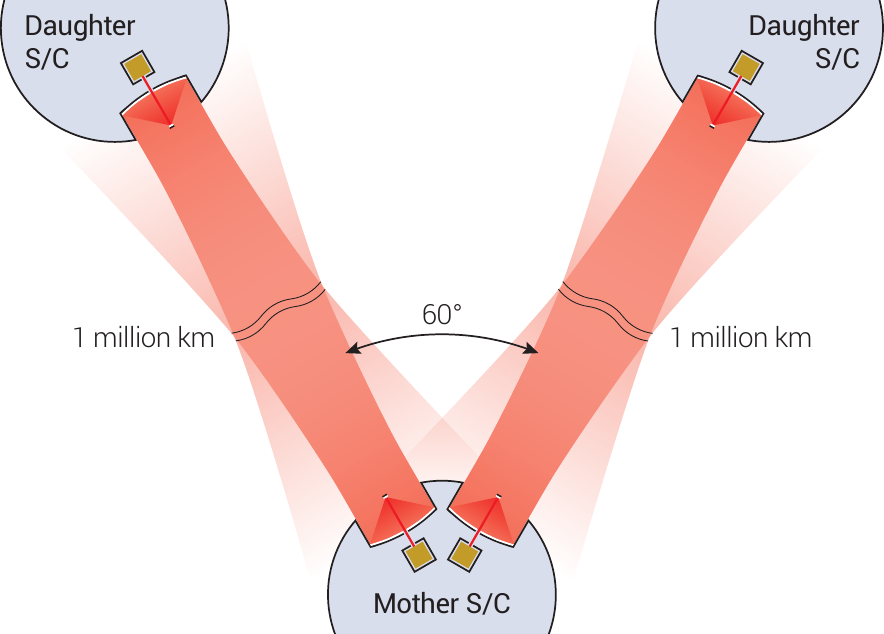}
\label{Fig:eLISAScheme}}
\hfil
\subfloat[The eLISA payload.]{\includegraphics[width=2.2in]{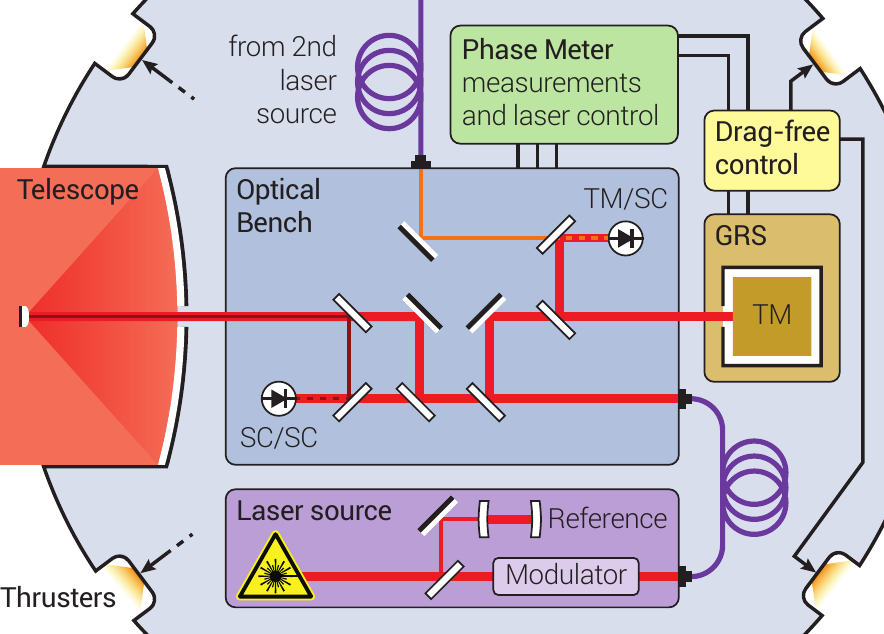}
\label{Fig:eLISAPayload}}

\caption{Measurement principle of eLISA}
\label{Fig:eLISAConfig}
\end{figure*}

To meet these requirements, the payload of eLISA consists of four identical units, two on the mother spacecraft and one on each daughter spacecraft.
Each unit contains a Gravitational Reference Sensor (GRS) with an embedded free-falling test mass that acts both as the end point of the optical length measurement and as a geodesic reference test particle. A telescope with 20 cm diameter transmits light from a 2W laser at 1064 nm along
the arm and also receives a small fraction of the light sent from the far spacecraft. Laser interferometry is performed on an optical bench in between the telescope and the GRS (see figure \ref{Fig:eLISAConfig}).

A major specificity of space-based interferometry between inertial references is the length mismatch of the arms and its time dependance. Actually, the arm length difference will be up to 30$\times10^3$~km with relative velocities (as seen from the 'mother' spacecraft) up to 7~m/s. As a consequence of the length mismatch, the different propagation delays along the two arms have to be taken into account in the post-processing algorithm to avoid the coupling of the laser frequency noise to the science signal. This algorithm is known as "Time Delay Interferometry" (TDI) \cite{Tinto2003, Shaddock2003, Shaddock2004, Tinto2005, Vallisneri2005, Vallisneri2007,Dhurandhar2009, Dhurandhar2013}. With a knowledge of the arm lengths at about 1~m accuracy, the laser source needs to be 'pre-stabilized' a a level of about $10^{-12}/\sqrt{\text{Hz}}$, using a ultra-stable Fabry-Perot cavity \cite{Thorpe2010}. 

Another method, called arm-locking has been suggested to further decrease the laser frequency noise. The arm-locking technique relies on the ultra-stable length of the eLISA arms in the 0.1 mHz to 10 Hz frequency range, by comparing the phase of the local laser with the phase of the incoming beams. The error signal from arm-locking is then used to tune the frequency of the laser source \cite{Sheard2003,Sylvestre2004,Sutton2008,Wand2009,McKenzie2009, Thorpe2011}.

The measurement of the inter-spacecraft distance is performed thanks to a low frequency modulation of the laser frequency, transmitting and correlating pseudo-random codes. Another modulation (at about 1~GHz) is also required to transfer the noise of the on-board clocks and compensate for their effect on the analog to digital converters (ADC) timestamping \cite{Esteban2009, Esteban2010, Sutton2010, Esteban2011, Heinzel2011, Otto2012, Sweeney2012, Sutton2013}.
These beat notes (carrying the 'science' as well as the ranging and clock synchronization signal) need to be measured at the level of the $\mu$cycle/$\sqrt{\text{Hz}}$, while following the frequency drifts between 5 and 25 MHz, due to the relative motion of the S/C.

Hence, the capability of eLISA to measure very small displacements relies (among other things) on accurate processing algorithms (TDI), precise feedback loops (arm-locking) and very low noise, extremely high performance instruments (phasemeters). Simulation software can simulate the Doppler effects, the propagation delays, reconstruction algorithms, etc (see e.g. \cite{Petiteau2008}). Nevertheless, the development of ÕhardwareÕ simulation are desirable, in order to characterize the detection devices, validate the numerical models and study the influence of the hardware on the detection algorithms. This is the purpose of the LOT (LISA On Table) experiment developed at the APC and described in the next sections.

Hardware simulators for eLISA have already been developed by other teams in the project. Two approaches have been implemented so far~: with realistic, electronicly delayed, phase noises \cite{Mitryk2010,Mitryk2012} and optical, short arms experiments \cite{deVine2010}. This paper describes an experimental setup allowing the simulation of optical links with the appropriate noise delays and Doppler shifts. The LOT experiment can reproduce most  of the features of the eLISA interferometry and expected noise. It will also be possible to inject user-defined noise function and accelerate the simulation to complete months of data simulation within days.

\section{Experimental setup}

\label{Sec:ExpSetup}
The goals of the LOT experiment is to be able to simulate optical beat notes, as representative as possible of the signals that will be recorded by eLISA. The experimental setup should also be kept very flexible to allow different configurations and the use of hardware prototypes, so that it could be adapted to new technologies or algorithms developed in the eLISA consortium.

Beyond the model of eLISA, the main challenge of such an experiment is to properly simulate delayed optical noise, while keeping the setup extremely stable on time scales of tens of seconds. The experimental setup presented here mainly focuses on the demonstration of the simulation principles and first results. The reduction of the interferometer intrinsic noises and complete development of eLISA functionalities will be addressed in future works.

\subsection{eLISA model}
\label{Subsec:Model}

\begin{figure}[h]
\centering
\includegraphics[width = \textwidth]{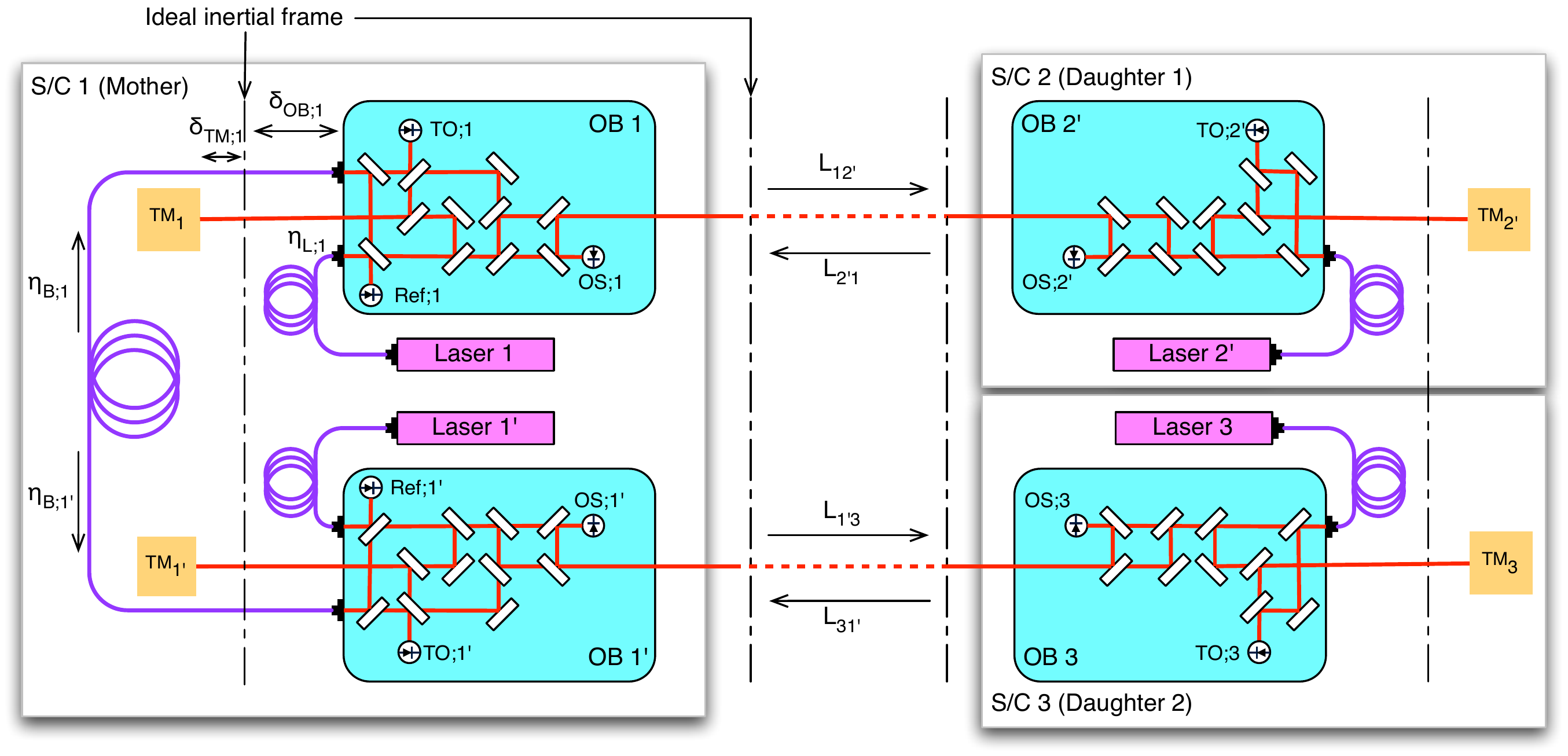}
\caption{Scheme of the intereferometirc measurements on the two arms of eLISA. See text for the explanations of the notations.}
\label{Fig:eLISAPrinc}
\end{figure}

The eLISA mission is composed of two arms, each of them linking one free-falling mass of the 'mother' spacecraft to the free-falling mass of each 'daughter' spacecraft. In practice, the variations of the arm length are computed from three interferometric measurements on each link~: test mass to optical bench on the mother S/C, 'mother' optical bench to 'daughter' optical bench, optical bench to test mass on the daughter S/C. The sum of these 3 measurements cancels out the (large) movement of the S/C w.r.t the test masses. The principle and different measurements of the eLISA interferometry is represented on figure~\ref{Fig:eLISAPrinc}.

eLISA consists of four (almost) identical payloads~: two on the mother S/C, one on each daughter S/C. Each payload consists of one test mass, one laser source and one optical bench (plus the control electronics, such as the phasemeter, the charge management unit, the housekeeping data unit, etc, which are not represented on figure~\ref{Fig:eLISAPrinc}). The three S/C are labelled 1, 2 and 3~: 1 refers to the mother S/C and the others to the daughters. Following conventions inherited from the previous LISA project (which had three identical S/C), devices and 'information' propagating clockwise are labelled with 1, 2 and 3, whereas they are labelled 1', 2' and 3' when propagating counter-clockwise.

On the mother optical bench 1, three interferometric measurements are performed, giving the relative optical phase between~: 
\begin{itemize}
\item the local laser (Laser 1) and the incoming signal from S/C 2 (and optical bench 2'), i.e. optical bench 1 to distant S/C~: $s_{OS;1}(t)$
\item the adjacent laser (Laser 1') and the local laser (Laser 1) after reflexion on the test mass 1 (TM${}_1$), i.e. test mass to optical bench signal~: $s_{TO;1}(t)$
\item the local laser and the adjacent laser, i.e. reference signal~:  $s_{Ref;1}$
\end{itemize}
The second optical bench on the mother S/C is strictly identical. A 'back-link' fiber transfers the light from bench 1 to bench 1' (and vice versa).
The optical benches on the daughter S/C are also identical except that there is only one optical bench (and one laser) per satellite, hence no reference beat note.

For the calculation, we will define the different positions noise w.r.t an ideal inertial frame, at the position of the S/C.
With the notations of the figure, the different signals (phases) recorded on the optical bench can be expressed as~:
\begin{subequations}
\begin{align}
s_{OS;1} &= \phi_{R;2'1}  - \delta_{OB;1} - \eta_{L;1} + o_{OS;1} \\
s_{TO;1} &= \eta_{L;1} + 2\times( \delta_{OB;1} + \delta_{TM;1} ) - \eta_{L;1'} - \eta_{B;1} + o_{TO;1} \\
s_{Ref;1} &= \eta_{L;1} - \eta_{L;1'} - \eta_{B;1} + o_{Ref;1} \\
\phi_{S;12'} &= \eta_{L;1} - \delta_{OB;1} + o_{S;12'},
\end{align}
\end{subequations}
 where $\phi_{R;2'1}$ is the phase of the laser beam received from optical bench 2'; $\delta_{OB;1}$ is the movement of the S/C 1 w.r.t the inertial frame (mainly due to thruster noise); $\eta_{L;1}$ is the phase noise of the laser source 1 (and $\eta_{L;1'}$ for the adjacent laser source); $\delta_{TM_1}$ is the movement of the test mass due to the residual acceleration noise; $\eta_{B;1}$ is the additional phase noise in the back link fiber from bench 1' to 1; $\phi_{S;12'}$ is the phase of the laser beam sent to optical bench 2' (w.r.t the inertial frame) and  $o_{xx;1}$ takes into account any local additional noise (such as the optical bench phase noise).
  
The same equations stand for the second optical bench of the mother S/C by replacing 1 with 1', 2 with 3' and 2' with 3. 
Similarly for optical bench 2', but with only one laser source~:
\begin{subequations}
\begin{align}
s_{OS;2'} &= \phi_{R;12'}  - \delta_{OB;2'} - \eta_{L;2'} + o_{OS;2'} \\
s_{TS;2'} &= 2\times( \delta_{OB;2'} + \delta_{TM;2'} ) + o_{TS;2'} \\
\phi_{S;2'1} &= \eta_{L;2'} - \delta_{OB;2'} + o_{S;2'1}
\end{align}
\end{subequations}

Moreover, the propagation along the arms between spacecraft leads to~:
\begin{align}
\phi_{R;2'1} &= D_{2'1} \phi_{S;2'1} + g_{w;2'1}  \nonumber  \\
&= D_{2'1} \eta_{L;2'} - D_{2'1} \delta_{OB;2'} + D_{2'1} o_{S;2'1} + g_{w;2'1},
\end{align}
with similar equations for $\phi_{R;31'}$, $\phi_{R;12'}$ and $\phi_{R;1'3}$. $D_{ij}$ is a delay operator~: $D_{ij} \phi (t) = \phi (t-L_{ij})$ ($L_{ij}$ being the light time from payload i to j, about 3.3~s); and $g_{w;ij}$ is the perturbation due to a gravitational wave. In eLISA, because of the relative motions of the S/C, $L_{ij} \neq L_{ji}$ (Sagnac effect) and $D_{ij} D_{kl} \phi \neq D_{kl} D_{ij} \phi$ (time evolving delays).

Combining these measurements gives an equivalent measure of the phase of the test mass position w.r.t to the incoming beam, corrected from the movement of the bench (w.r.t the test mass)~:
\begin{subequations}
\begin{align}
s_{TT;1} &= s_{OS;1} + \frac{s_{TO;1}+s_{Ref;1}}{2} + D_{2'1} \frac{s_{TS;2'}}{2} \nonumber \\
&= D_{2'1} \eta_{L;2'} -  \eta_{L;1} + g_{w;2'1}    + \delta_{TM;1} + D_{2'1} \delta_{TM;2'} + o_{TS;1} + D_{2'1} o_{S;2'1} \\
s_{TT;1'} &= s_{OS;1'} + \frac{s_{TO;1'}+s_{Ref;1'}}{2} + D_{31'} \frac{s_{TS;3}}{2} \nonumber\\
&= D_{31'} \eta_{L;3}  - \eta_{L;1'} + g_{w;31'}   +  \delta_{TM;1'} +D_{31'} \delta_{TM;3}  + o_{TS;1'} + D_{31'} o_{S;31'} \\
s_{TT;2'} &= s_{OS;2'} + \frac{s_{TO;2'}}{2} + D_{12'} \frac{s_{TS;1}}{2} \nonumber \\
&=D_{12'} \eta_{L;1} - \eta_{L;2'} + g_{w;12'}  +  \delta_{TM;2'} + D_{12'} \delta_{TM;1} + o_{TS;2'} + D_{12'} o_{S;12'}\\
s_{TT;3} &= s_{OS;3} + \frac{s_{TO;3}}{2} + D_{1'3} \frac{s_{TS;1'}}{2} \nonumber \\
&= D_{1'3} \eta_{L;1'} - \eta_{L;3} + g_{w;1'3}  + \delta_{TM;3} +  D_{1'3} \delta_{TM;1'} +  o_{TS;3} + D_{1'3} o_{S;1'3}
\end{align}
\label{Eq:TTequations}
\end{subequations}

These equations would not allow any differential interferometric measurement between the two arms without the knowledge of the relative phase of laser sources 1 and 1' (i.e at the vertex of the interferometer). The fiber link from bench 1 to 1' (and vice-versa) adds a noise level that can be far above the requirements. However, the fiber is \emph{reciprocal} \cite{Fleddermann2009}, which means that the added noise from 1 to 1' is identical to the noise added when propagating from 1' to 1~: $\eta_{B;1} = \eta_{B;1'}$. In that case, the differential noise between laser 1 and 1' can be deduced from $s_{Ref;1}$ and $s_{Ref;1'}$~:
\begin{equation}
\frac{s_{Ref;1} - s_{Ref;1'}}{2} = \eta_{L;1} - \eta_{L;1'} + \frac{o_{Ref;1}-o_{Ref;1'}}{2}
\label{Eq:BacklinkEq}
\end{equation} 

\begin{figure*}[h]
\centering
\subfloat[eLISA noise spectrum (averaged strain sensitivity)]{\includegraphics[width=0.8\textwidth]{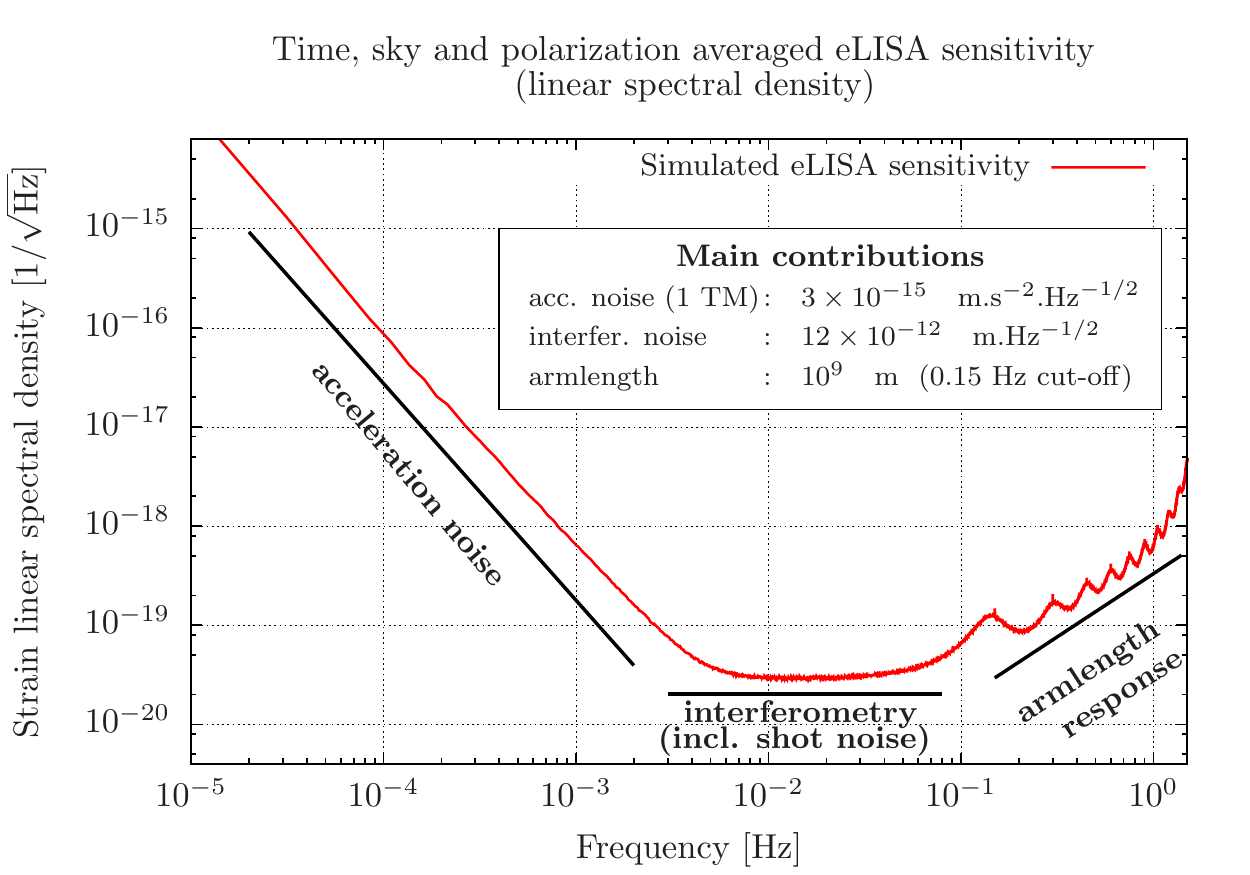}
\label{Fig:eLISANoiseSpec}}
\hfil
\subfloat[Examples of GW astrophysical sources in the frequency range of eLISA, compared with the sensitivity curve of eLISA (for more details on its computation, see \cite{ELISAWP2013} an references therein)]{\includegraphics[width=0.8\textwidth]{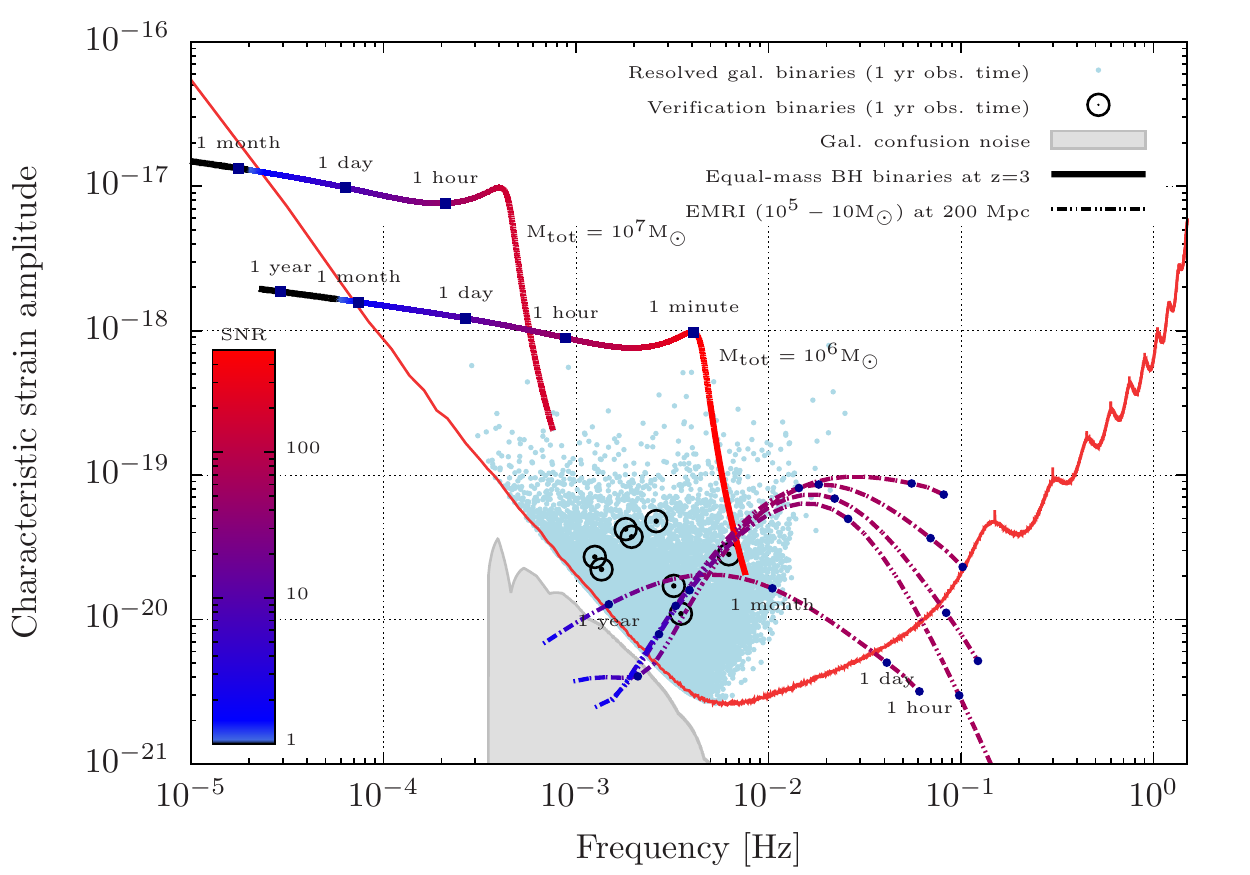}
\label{Fig:eLISASensCurve}}

\caption{Noise spectrum and associated eLISA sensitivity.}
\label{Fig:eLISAPerformance}
\end{figure*}

Equations \ref{Eq:TTequations} and \ref{Eq:BacklinkEq} form the set of time-delayed interferometry. The dominant noise terms are, by many orders of magnitudes, the laser phase fluctuations. These noises are however transported from one S/C to another and appear as differences between the local laser and the distant, delayed one. Their contribution can be canceled by properly delaying and combining the $S_{TT;q}$ signals~: this method is known as  time delay interferometry (TDI) (see e.g. \cite{Tinto2005}). 
TDI however requires a precise knowledge of the light time between S/C (at the meter level over $10^6$~km). Moreover, these equations assumed perfect clocks on each S/C, with no relative drifts nor jitter. Ranging and clock noise transfer can be performed by adding auxiliary modulations on the laser links \cite{Heinzel2011}. After TDI and clock noise corrections \cite{Otto2012}, the residual phase noise should be due to the test masses (residual acceleration noise at low frequency) and the optical noise (including shot noise and path length fluctuations on the optical benches). Convolved with the eLISA response to the gravitational waves (including the TDI response function), the expected noise curve and associated sensitivity curve are represented on figure~\ref{Fig:eLISAPerformance} (for more details on the sensitivity computation and science objectives of eLISA, refer to \cite{ELISAWP2013, eLISAWebSite2013, NGOYB2012} and references therein).

The primary goal of the LOT experiment is to optically simulate the $S_{TT;q}$ signal and then apply the TDI algorithm on the recorded signals, but it can also be configured to simulate any of the $s_{XX,q}$ signal.

Practically, the laser frequencies will not be let freely running but will be phase locked on a master, frequency stabilized source (e.g. Laser 1). 
From the previous equations, assuming perfect correction, this phase locking means that~:
\begin{subequations}
\begin{align}
s_{Ref;1} &= 0 \Rightarrow \eta_{L;1'} = \eta_{L_1} - \eta_{B;1} + o_{Ref;1}\\
s_{OS;2'} &= 0 \Rightarrow \eta_{L;2'} = D_{12'} \eta_{L;1} + g_{w;12'} - D_{12'} \delta_{OB;1} -  \delta_{OB;2'} + D_{12'} o_{S;12'} + o_{OS;2'} \\
s_{OS;3} &= 0 \Rightarrow \eta_{L;3} = D_{1'3} \eta_{L;1'} + g_{w;1'3} - D_{1'3} \delta_{OB;1'} -  \delta_{OB;3} + D_{1'3} o_{S;1'3} + o_{OS;3} 
\end{align}
\end{subequations}

Assuming that $\delta_{OB;q}=0$ and $\eta_{B;q}=0$ (these noises can be subtracted using $s_{TO;q}$ and $s_{Ref;q}$ signals) and neglecting both test mass and other local noises ($\delta_{TM} = o_{xx;q} = 0$), this configuration is effectively equivalent to a transponder, where the phase noises of the incoming beams on S/C 2 and 3 are transferred on the beams sent back to S/C 1.

With these assumptions, the 'test-mass to test mass' (equations \ref{Eq:TTequations}) signals become~:
\begin{subequations}
\begin{align}
s_{TT;1} &= D_{2'1}D_{12'} \eta_{L;1} -  \eta_{L;1} + D_{2'1} g_{w;12'}   + g_{w;2'1}  \\
s_{TT;1'} &= D_{31'}D_{1'3} \eta_{L;1} - \eta_{L;1} + D_{31'}g_{w;1'3}  + g_{w;31'}  \\
s_{TT;2'} &= 0\\
s_{TT;3} &= 0
\end{align}
\label{Eq:TTequationsTransp}
\end{subequations}

These simplified, 'transponder' equations (which require only two simulated data streams), have been used for the first experimental measurements performed on the LOT, with the additional constraint of a static constellation (i.e $D_{2'1}D_{12'}$ and $D_{31'}D_{1'3}$ are constant and commute) and no gravitational wave signal.
In that case, the simplest TDI combination (X, first generation) leads to~:
\begin{equation}
X_{1^{st}} = (1 - D_{31'}D_{1'3})s_{TT;1} - (1 - D_{2'1}D_{1'2})s_{TT;1'}
\label{Eq:XFirst}
\end{equation}
This combination synthesized an equivalent Michelson interferometer with equal arms.

\subsection{Experimental setup}
\label{Subsec:ExpSetup}

\paragraph{Overview}
The goal of the eLisa On Table (LOT) experiment is to make an experimental approach of the main subjects of the eLISA mission such as phase noise or TDI (time delay interferometry). It is innovative insofar as it includes both electronic and optical aspects of a hardware simulator in one experiment and allows to simulate optical beatnotes with appropriate phase delays.
Furthermore, as it is entirely computer controlled, the implementation of realistic noises  and delays can be done through the use of software simulating the mission, such as LISAcode\cite{PetiteauThese2008}. The LOT is expected to be a very complete optical simulator representative of eLISA, offering the additional flexibility to define ad-hoc fluctuations in the frequency, amplitude of phase of the laser and propagation delays. The control electronics also allows to 'accelerate' the simulation, so that one year of data streams could be generated within a few days.\\

The optical part of the LOT is mainly based on a Mach-Zehnder interferometer combined to AOMs (acousto-optic modulator) to shift the laser frequency on the arms in order to obtain heterodyne interferences. The beatnotes are measured by photodiodes and sent to a phasemeter. Simultaneously, the simulated RF commands are electronicly mixed, low-pass filtered, and sent to other channels of the phasemeter. 
Both, the AOMs and the phasemeter are controlled by a labview program with predefined mathematics models (computed by a separate C library for maximum efficiency) used to generate the RF frequencies and the simulated noise used by the AOMs, including user-defined delays. These delays are held fixed in the present work, but will be representative of the eLISA orbits once the command program is combined with LISAcode \cite{Petiteau2008,PetiteauThese2008}.\\

\begin{figure}[h]
 \includegraphics[trim = 110mm 40mm 55mm 20mm, clip, width=\textwidth]{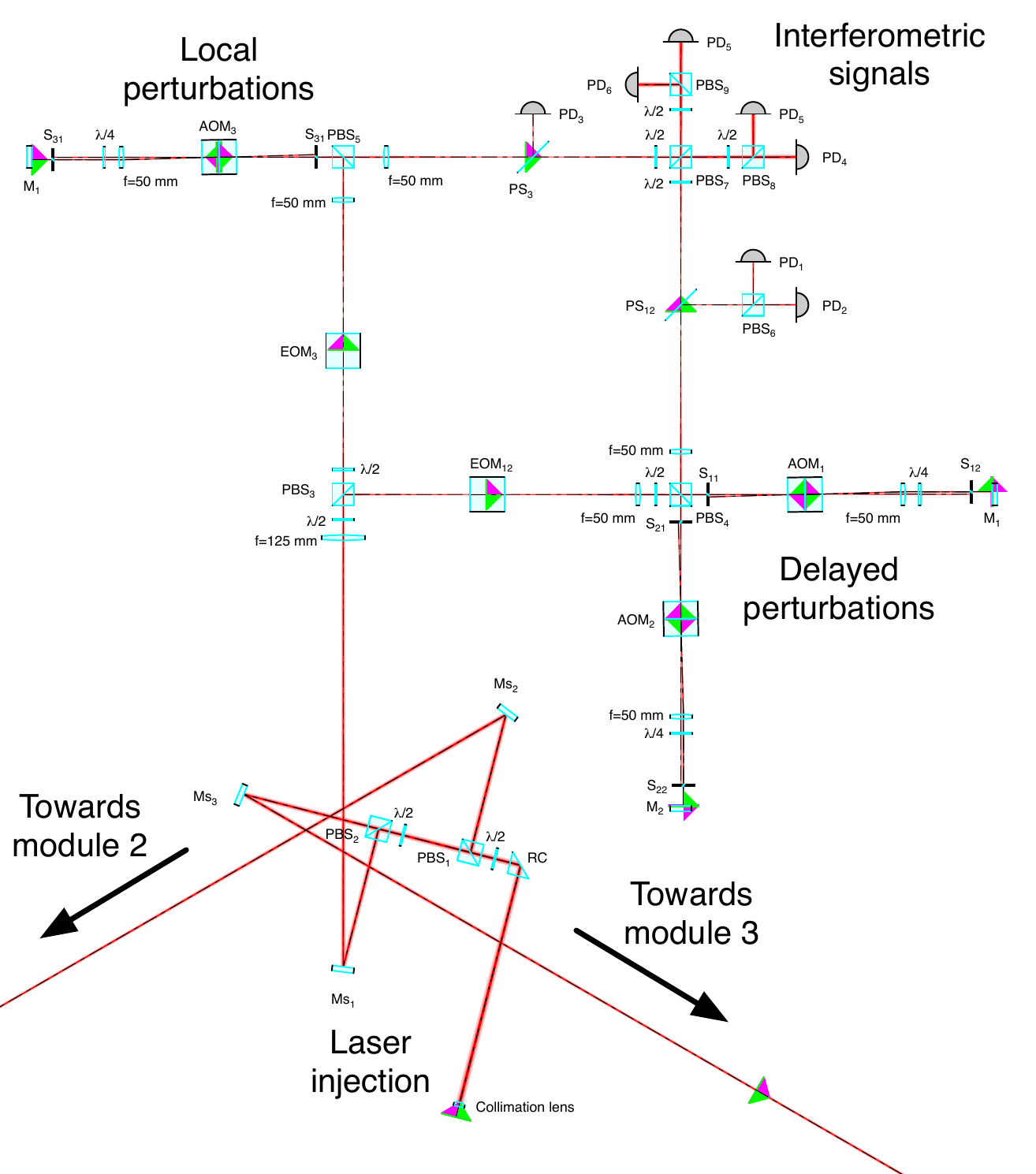}
 \caption{Optical scheme of one module, representative of the mother S/C, simulated with OptoCad \cite{OptoCAD2013}.}
 \label{Fig:OpticalLayoutOneModule}
\end{figure}

\paragraph{Description of the optical setup}
The LOT's optical part is shown on Fig.\ref{Fig:OpticalLayoutOneModule}. which represents one module of the simulator. The LOT is composed of two of those modules, each one represents one satellite of the eLISA configuration. The third module, representative of the third satellite, will be implemented in future works. For this study only one module has been used, representative of the mother S/C.

\begin{figure}[h]
 \includegraphics[trim = 113mm 155mm 131mm 35mm, clip, width=\textwidth]{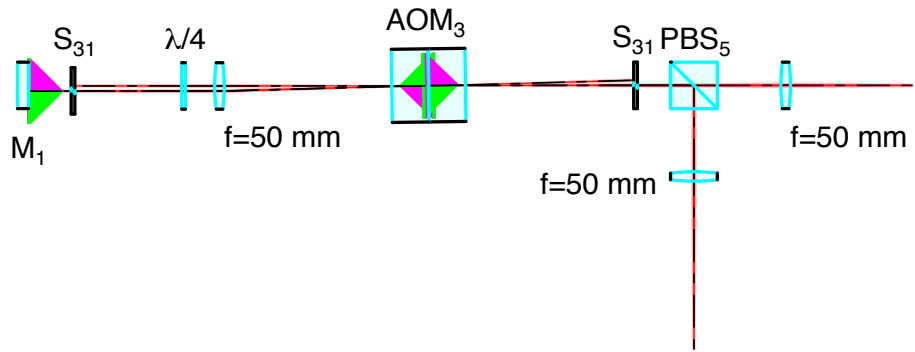}
 \caption{Optical scheme of cat's eye configuration for one arm}
 \label{cat}
\end{figure}

As for eLISA, the module representing one satellite has three interfering beams, one 'local' and two 'distant'. A single laser source at 1064–m (Innolight Mephisto 500) is used to produce these beams using a combination of polarizing beam splitters and waveplates so that optical paths of the distant beams follow the same optical path but with orthogonal polarizations. The local arm's frequency can be shifted with AOM 1. In the same way, AOM 2 and 3 induce the frequency shift for the two distant arms. 
Each of these distant arms interferes with the local arm to produce a heterodyne signal detected by 48 MHz bandwith photodiodes with power noise below 8 pW/$\sqrt{Hz}$.
Free space A\&A MT110-B50 TeO2 AOMs have been used for the LOT. The maximum diffraction efficiency ($\ge 80$\%) is reached at 110 MHz, with a bandwidth of $\pm 15$~MHz with an efficiency greater than 65\%. This broad modulation bandwidth is particularly useful for simulations of the doppler effect. 
However, since a large frequency shift induces also a large angular shift after the AOM, a cat's eye configuration has been implemented for the frequency modulation (see figure \ref{cat})  

In such a configuration, the s-polarized laser beam is sent to the AOM using a polarizing beamsplitter ($PBS_5$). 
After passing through the AOM ($AOM_3$), two beams are present~: the order 0 is the unaffected beam, while order 1 is frequency shifted and deflected, proportionally to the RF frequency sent to the AOM. The diffraction efficiency is also a function of the RF power. After this first pass trough the AOM, the order 0 beam is blocked using a small slit ($S_{31}$).

Using a combination of lens (with its focal point at the center of the AOM), mirror and $\lambda$/4 waveplate, the order 1 beam is sent back on the same path with an orthogonal polarization, whatever the deflection angle (i.e. the RF frequency).When passing through the AOM, the return beam is partially frequency and angular shifted. Another slit is used to only select the doubly shifted beam, which is then p-polarized and goes straight trough the beam splitter.

In the present experiment, RF commands around 110 and 106 MHz are sent, respectively to the two distant and to the local AOMs, producing two optical beat signals at about 8 MHz. The command frequency sent to the distant AOMS are slightly different to produce beatnotes separated by a few kHz, in order to avoid parasitic mixing  from the polarization leakage of the optical components.

\paragraph{Description of the electronic interferometer and command-control apparatus}

\begin{figure}[h]
\includegraphics[width=\textwidth]{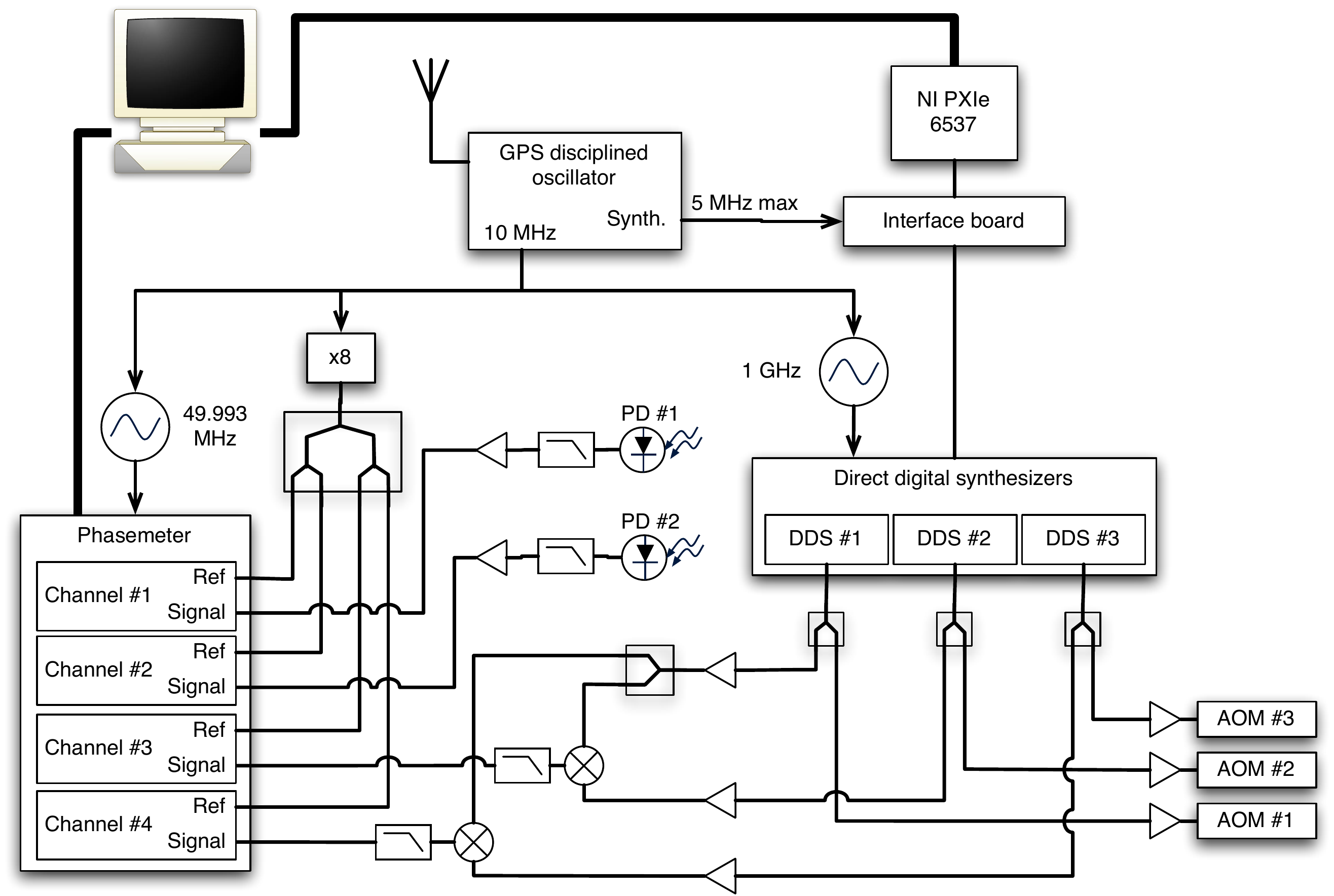}
 \caption{Electronic scheme of the LOT}
 \label{Fig:ElectronicalLayoiut}
\end{figure}

The LOT's electronic part is composed of all necessary devices for control and measurement but also of an electronic version of the LOT which recovers and combines the signals sent to the AOMs in order to analyze and compare them to the optical signal. The concept is illustrated on Fig.\ref{Fig:ElectronicalLayoiut}.

The data streams for each channel (i.e. sent to the AOMs) are simulated (i.e. generated, delayed and interpolated), converted to digital commands and sent to a National Instrument PXie~6537 communication board. The frames (being 264 bits long) are serially sent on some of the 32 output ports of the NI card. the communication rate is controlled using a synthesized clock, up to 5 MHz, derived from the GPS disciplined oscillator. The communication card as well as the DSS can handle communication rates up to 50~MHz (i.e. a frames' rate of about 190 kHz), while it was set to 2~MHz in the present work (frames rate of 7.6~kHz).

Computer controlled DDS (Direct Digitial Synthesizer, model Agilent AD9912) actuates the AOMs. 
The signal of each DDS channel is split, amplified and mixed before going through a low-pass filter and being measured by the phasemeter. DDS channel 1 represents the local arm, channels 2 and 3 stand for the two distant arms. Just as the local laser shifted by AOM \#1 interferes respectively with the distant lasers shifted by AOM \#2 and \#3, the signal generated by DDS \#1 is mixed respectively with the DDS channels \#2 and \#3 before being sent to phasemeter channels 3 and 4. 
The DDS are able to generate RF signal up to 400 MHz with an accuracy of 3.6 $\mu$Hz and the possibility to adjust the phase of the signal with a precision of 0.38 mrad. The electronics clocks are derived from a 10~MHz high stability, GPS disciplined oscillator to reduce the differential jitter noise. 

The phasemeter used in this study is a prototype developed by the Albert Einstein Institut in Hanover, Germany \cite{Bykov2009}. Within the phasemeters, the signals are digitized using an ADC (Analog to Digital Converter), running at the phasemeter reference clock (presently 49.993~MHz). The frequency measurements are then performed by monitoring the frequency of a digital phase locked loop. In order to compensate for the jitter noise of the input ADC, each channel is combined with a 80~MHz reference signal, produced from the octupled 10~MHz GPS frequency. This signal (folded at 20 MHz after sampling) is processed in the same way as the 'science' data  and used to correct it. A residual noise lower than 1$\mu$Hz$\sqrt{\text{Hz}}$, for input frequencies between 2 and 20 MHz can be achieved. The output data are transmitted to the computer using a parallel port, at a rate of about 23.8~Hz.

All the experiment is performed in a clean room. The optical table is placed on an air cushion to reduce high frequency noise and all optical devices put under a box in order to reduce noises induced by air flow. Also, a heat device is fixed on the top to implement temperature layers so that air perturbation induced by eventual warm spots on the table are quickly absorbed.\\

One of the main applications of the LOT will be to test the TDI procedure by injecting noise into the arms through the AOMs with different delays for each arm, and apply the TDI algorithm on the recorded phase data. The first characterization of the LOT experiment and TDI measurements are the subject of the next section.

\section{First experimental results}

Different configurations have been used to perform the first tests on the experimental setup described above. For all the measurements, the communication clock (synthesized from the GPS oscillator) was set to 2~MHz, meaning an update of the DDS data at a rate of 7.6~kHz. Given the theoretical DDS resolution of 3.6$\mu$Hz, such a frequency leads to a quantification noise of about $3\times 10^{-8}$~Hz/$\sqrt{\text{Hz}}$. 

Unless otherwise noted in the following descriptions, the carrier frequencies were set to 108.355~MHz ('local' arm), 112.5~MHz (Distant arm 1) and 112.6~MHz (Distant arm 2). This values lead to electronic beat notes around 4.15 and 4.25 MHz. The optical beat notes are twice these values (due to the double pass within the AOMs).

The frequency modulations of these carrier frequencies were simulated as gaussian white noises of amplitude 140~Hz/$\sqrt{\text{Hz}}$ (mono lateral amplitude spectral density) over a bandwidth of 500~Hz. This simulated signal is delayed (according to the simulation parameters) and up-sampled to 7.6 kHz using a $7^{th}$~order Lagrange interpolation filter. No amplitude, nor phase modulation were sent to the AOMs.

The delays were held constant, at 6.5020814914~s (arm 12) and 6.7118260556~s (arm 13). These values corresponds to typical round-trip delays on eLISA and, to avoid the influence of post-processing interpolation filters, they correspond to an integer number of phasemeter sampling periods (155 and 160 respectively).

Four streams of data are recorded by the four-channel phasemeter (ordered from channel 1 to 4)~: $s_{o,1}$, $s_{o,1'}$, $s_{e,1}$, $s_{e,1'}$, where $s_{o,q}$ refers to the 'optical' interferences and  $s_{e,q}$ to the 'electronic' ones. 

The amplitude spectral density (ASD) of the signals (raw or reconstructed) are computed on a logarithmic frequency axis using the algorithm described in \cite{Trobs2006,Trobs2009}.

\subsection{Intrinsic noise of the simulator}

Three experiments have been conducted to estimate the intrinsic noise of the system and the analysis process.

\begin{figure}[h]
\includegraphics[width=\textwidth]{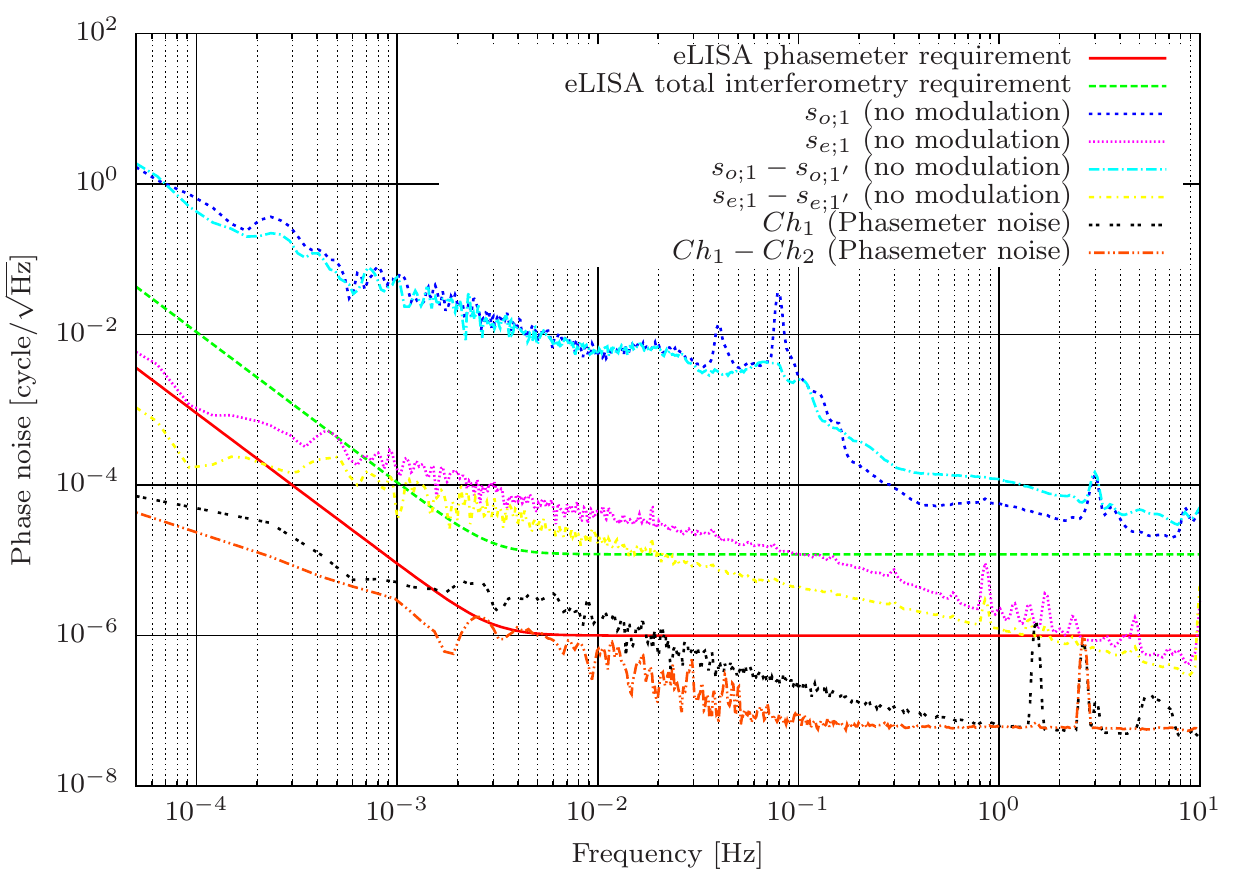}
\caption{Reference levels of noise in the LOT experiment~: phasemeter noise, DDS and electronic noise, optical noise}
\label{Fig:IntrinsicNoise}
\end{figure}

First, the intrinsic noise level of the phasemeter has been estimated using a single RF source at 2.001~MHz, split on the 4 input channels of the phasemeter. The  ASD of data recorded on channel 1 and the ASD of the difference between channel 1 and 2 are represented on figure~\ref{Fig:IntrinsicNoise}. The ASD of raw values and differences between other channels give very similar results. These results show that the raw data are slightly above the eLISA requirements for the phase measurement noise, while the differential measurement between two channels is marginally compatible with the requirement. The difference between the two curves are due to a relatively strong common mode between the channels, whose origin is unclear for the moment, but could be due, e.g. to a residual phase jitter between the reference signal (at 72.001 MHz in this experiment) and the synthesized 2.001~MHz signal.  However, the intrinsic phase noise of the phasemeter is well below the other noise sources of the LOT experiment and will not be a concern in the present work. For reference, the requirements for the interferometric measurement (including shot noise, optical bench noise, electronics, etc.) are also plotted on figure~\ref{Fig:IntrinsicNoise}.

On a second configuration, the 3 DDS were configured with no modulation, and carrier frequencies at 108.355 MHz ('local' arm), 112.572 MHz ('distant' arm 1) and 112.583~MHz ('distant' arm 2). The ASD of the resulting beat notes $s_{o,1}$, $s_{e,1}$  and the combinations $s_{o,1}$ - $s_{o,1'}$ , $s_{e,1}$ - $s_{e,1'}$ are represented on figure~\ref{Fig:IntrinsicNoise} ($s_{o,1'}$ and $s_{e,1'}$ have similar spectra as, repectively, $s_{o,1}$ and $s_{e,1}$). As expected, the 'optical' beat notes are dominated by optical pathlengths fluctuations, due to thermal expansion of the aluminum support plates, the air turbulences, etc.  The two optical signals are mostly uncorrelated except for the 2 resonant peaks at 0.04 and 0.08 Hz (maybe related to the air damping of the optical table).

The noise associated with the DDS (in this configuration) is about one order of magnitude below the optical noise. The differential measurement is noticeably lower than the raw signal, showing a correlation of the jitter between the different DDS. It could be due to a residual jitter noise between the 1 GHz synthesized signal (the DDS clock reference) and the RF reference signal ay 80~MHz (from the octupled GPS frequency).

Applied to the present configuration, the 'Michelson' $X_{1^{st}}$ combination (see eq~\ref{Eq:XFirst}) leads to~:
\begin{equation}
X_{1^{st};x}(t) =  s_{x,1}(t) - s_{x,1}(t-\tau_3) - \left[ s_{x,1'}(t) - s_{x,1'}(t-\tau_2) \right],
\label{Eq:XTDIStatic}
\end{equation}
where $\tau_3$ is the simulated round-trip time between S/C 1 and 3 ( 6.711826~s) and $\tau_2$ is the simulated round-trip time between S/C 1 and 2 (6.502081~s).

In the case of uncorrelated noise $s_{x,1}$ and $s_{x,1'}$ the amplitude spectral density of $X_{1^{st};x} $ is given by~:
\begin{subequations}
\begin{align}
\widetilde{X_{1^{st};x}}(\nu) &= \sqrt{\widetilde{s_{x,1}}^2(\nu) \sin^2( \pi \tau_3 \nu) + \widetilde{s_{x,1'}}^2(\nu) \sin^2( \pi \tau_3 \nu)}  \label{Eq:TFuncXFull}\\
&=\widetilde{s_{x,1}}(\nu) \times \sqrt{1-\cos(\pi (\tau_2+\tau_3)\nu)\cos(\pi (\tau_2-\tau_3)\nu)} \label{Eq:TFuncXSimp},
\end{align}
\end{subequations}
where equation \ref{Eq:TFuncXSimp} assumes similar ASD for the $s_{x;q}$ signals ($\widetilde{s_{x,1}} = \widetilde{s_{x,1'}}$). With the same assumption and at low frequencies, we have $\widetilde{X_{1^{st};x}}(\nu) \approx \pi \sqrt{\tau_2^2 + \tau_3^2} \times \nu \widetilde{s_{x,1}}(\nu)$~: $X_{1^{st}}$ has a linear response function for $\nu \ll 2/(\tau_2+\tau_3)$.

\begin{figure}[h]
\includegraphics[width=\textwidth]{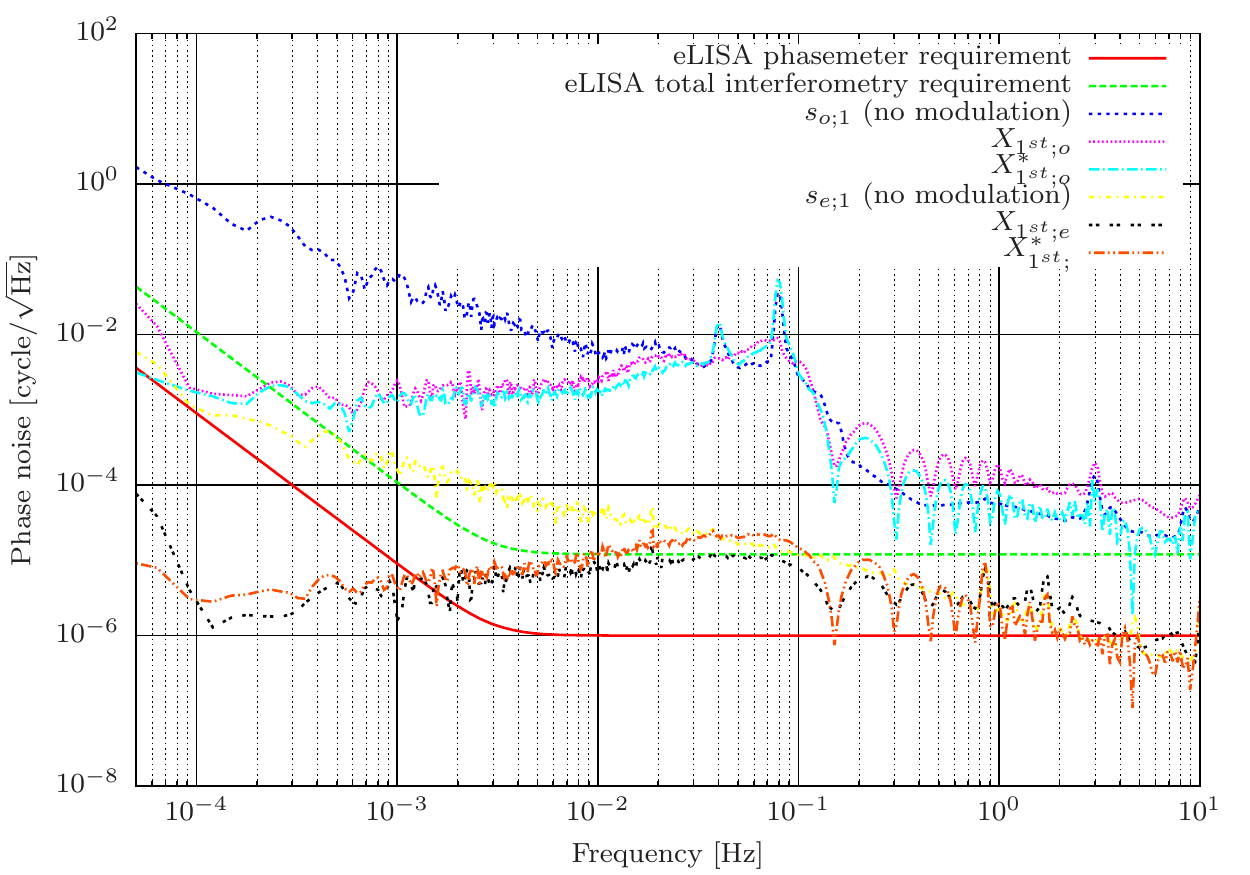}
\caption{Reference levels of noise in the LOT experiment before and after application of TDI $X_{1^{st}}$ on not modulated signals and comparison with the theoretical transfer function  $X^{{}*{}}_{1^{st}}$}
\label{Fig:TestTDINoModulation}
\end{figure}

This approximation was tested on the recorded signal (with no modulation) and plotted on figure~\ref{Fig:TestTDINoModulation}.

The results obtained after application of TDI and its theoretical model (transfer function) show identical behavior, especially the position of the nodes and the noise level. The observed discrepancies could be due to the correlation between the recorded signals.  These noise levels are expected to be the noise floor of the present configuration of the LOT, provided a perfect cancellation of the simulated frequency modulations applied on the different AOMs. 

\begin{figure}[h]
\includegraphics[width=\textwidth]{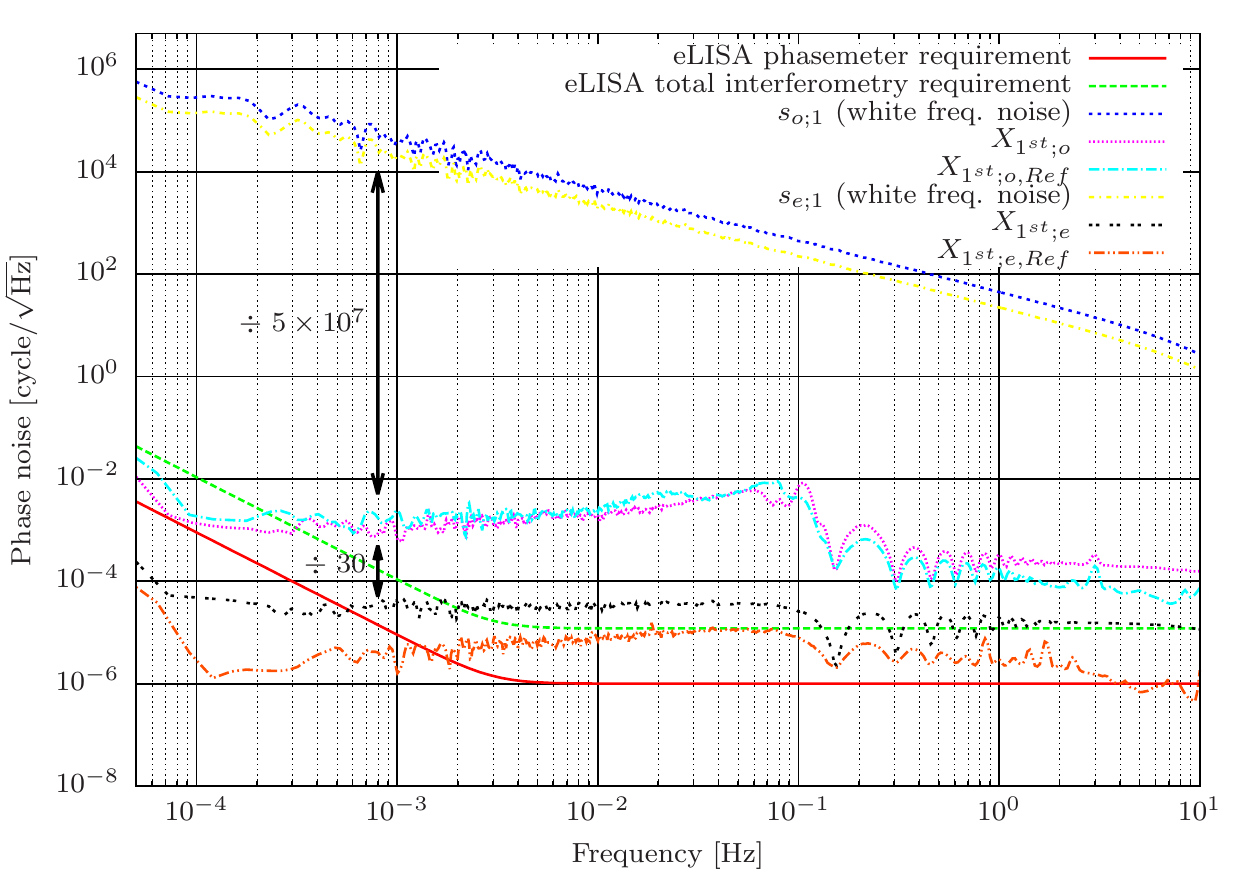}
\caption{Reference levels of noise in the LOT experiment before and after application of TDI $X_{1^{st}}$ on identical, noisy signals and comparison with the intrinsic noise level of the experiment (see fig.~\ref{Fig:TestTDINoModulation}).}
\label{Fig:TestTDIWhiteNoise}
\end{figure}

As a third test of the intrinsic noise of the experiment, the same frequency modulation (white frequency noise) has been applied on the two distant arms. This white noise (at a level of $280$~Hz/$\sqrt{\text{Hz}}$) is representative of the  expected frequency noise of the stabilized laser source.
The data were processed as described above, applying TDI with the same delay (6.502081~s) on the two arms. The ASD of the raw and processed signals are represented on figure~\ref{Fig:TestTDIWhiteNoise}, together with the estimated intrinsic noise level of the LOT experiment. The residual optical noise is compatible with the reference level of the previous configuration, except above about 1~Hz. The TDI combination of the electronically signals is about 10 times above its reference level. This discrepancy will be investigated in future works, but could be due to a small offset between the timestamps of the different channels, effectively equivalent to a differential delay between the signals. Above 1 Hz, the discrepancy seems to be correlated with the frequency cut-off of the phasemeter (possible differential phase or amplitude variations).

However, these first analyses show a noise reduction factor of $5\times 10^7$ on the optical signals and about $10^9$ on the electronic signals at 1~mHz. On eLISA a reduction factor of $10^9$ to $10^{10}$ is expected.

\subsection{Measurements with delayed perturbations}

To perform a first test of TDI with the LOT in a realistic configuration, a white noise (amplitude $280$~Hz/$\sqrt{\text{Hz}}$) was simulated and applied on the local AOM (no delay), AOM 2 and AOM 3. This setup is representative of a static 'transponder' eLISA. The signal sent to the local AOM is not delayed, whereas the signal on AOM 2 is delayed by 6.502081~s and AOM 3 by 6.502081~s (equal arm lengths configuration) or 6.711826~s (unequal arm lengths).

\begin{figure}[h]
\includegraphics[width=\textwidth]{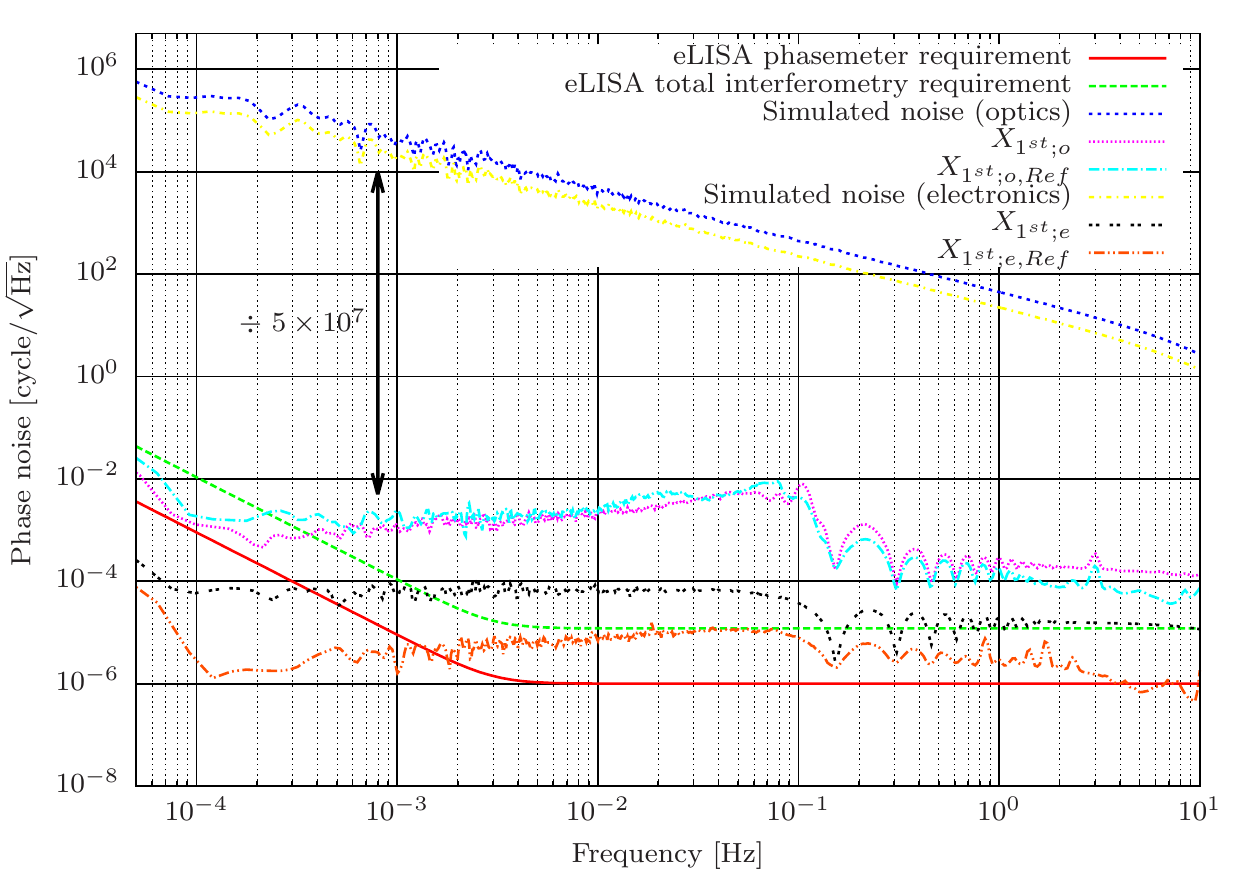}
\caption{Reference levels of noise in the LOT experiment before and after application of TDI $X_{1^{st}}$ on noisy signals with identical delays (6.502081~s) and comparison with the intrinsic noise level of the experiment (see fig.~\ref{Fig:TestTDINoModulation}).}
\label{Fig:WhiteNoiseTDIEqualLengthLOT}
\end{figure}

\begin{figure}[h]
\includegraphics[width=\textwidth]{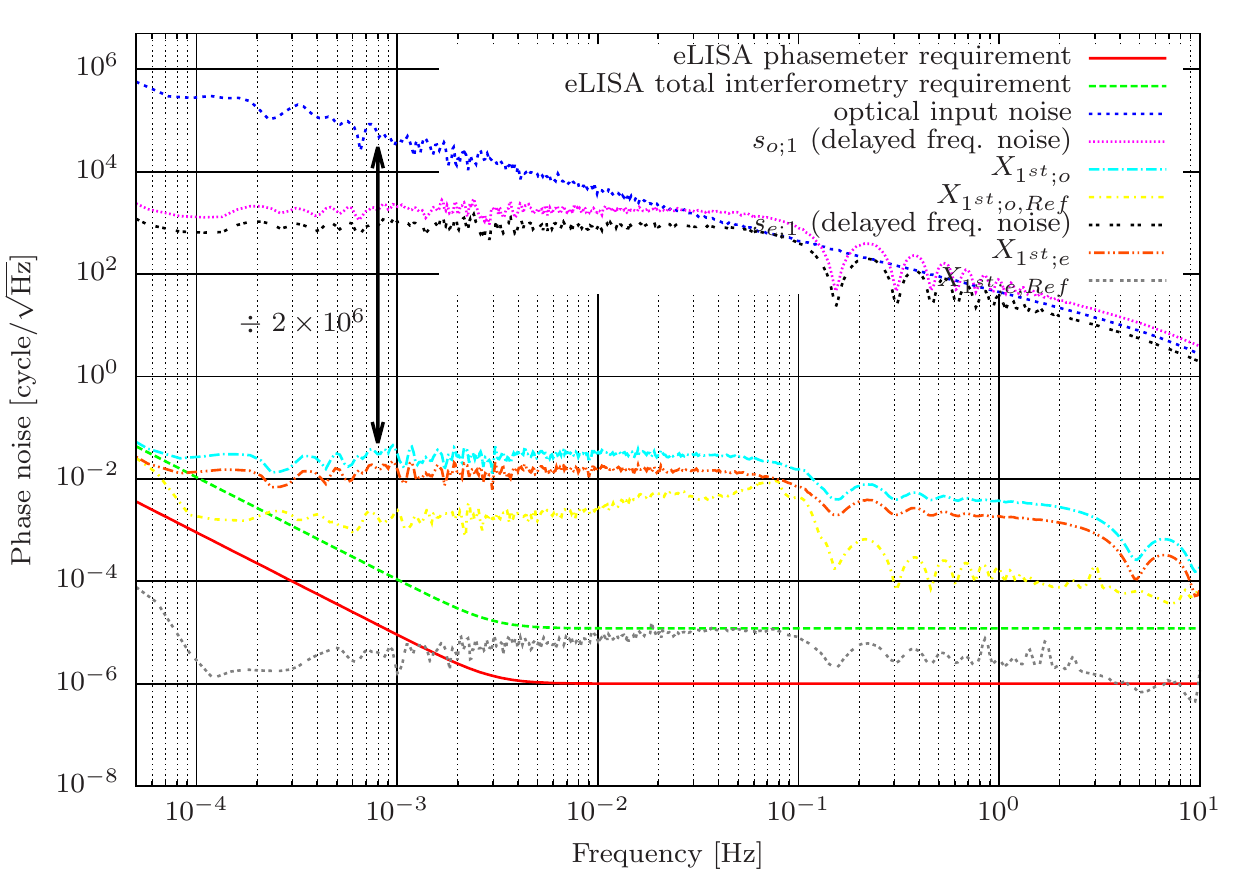}
\caption{Reference levels of noise in the LOT experiment before and after application of TDI $X_{1^{st}}$ on delayed, noisy signals and comparison with the intrinsic noise level of the experiment (see fig.~\ref{Fig:TestTDINoModulation}).}
\label{Fig:WhiteNoiseTDILOT}
\end{figure}

The same analysis as described above is performed, using the TDI combination given in equation~\ref{Eq:XTDIStatic} and the results are given on figure~\ref{Fig:WhiteNoiseTDIEqualLengthLOT} (identical delays) and \ref{Fig:WhiteNoiseTDILOT} (unequal arm lengths configuration). 

The results for the equal arm length configuration is very similar to figure~\ref{Fig:TestTDINoModulation}, which differs from this experiment due to the simulation of local noise and delayed 'distant' noise, therefore inducing a different noise spectra of the recorded beat notes.  the only noticeable effect is on the level of the residual noise for the electronic interferometer ($X_{1^{st};e}$), which is mainly due to the relative jitter between the output signals of the DDS.

When simulating unequal arm-lengths, the residual noise level increases by about 2 orders of magnitude, for a noise reduction factor of only $2\times 10^6$, on the optical and electronic channels. The node at about 5~Hz ($\approx 1/(\tau_3-\tau_2)$) suggests the effect of time jitter between the phasemeter clock (timestamps) and the actual update rate of the DDS frequencies (taking into account the GPS synthesized clock, the interface board and the PXIe controller). This effect can mimic fluctuations of the arm-length and the required accuracy on the propagation time should be of the order of 10~ns (i.e. 3~m) to achieve the expected noise cancellation.

\section{Next steps}

The results of the first measurements, as described above, demonstrate the validity of the concept but also clearly show evidence of two major sources of noise~: jitter in the simulated delays and optical path-length noise.
The next steps and future works will address these two points and also increase the representativity of the simulation.

First, additional measurements will be needed to clearly identify the origin of the residual noise, but some developments are already undertaken or foreseen to improve there performances of the LOT experiment.

The time jitter of the simulated delays will be addressed by inserting an FPGA (Field Programmable Gate Array) board right after the NI PXIe 6537 communication board. This FPGA will take charge of buffering, delaying and synchronizing the command frames to the AOMs. It will also allow to add frequency corrections, taken from digital or analog error signals. With the appropriate delays, this capability will mimic the phase lock of the laser sources on the distant S/C, as well as the implementation of the arm-locking stabilization scheme.This FPGA board is currently under development. 
Additionally, a ranging pilot tone (frequency modulation around 1~Hz) will be implemented to monitor the effective delay of the frames (this technique has already been successfully used in \cite{Mitryk2012}).

The optical bench phase noise will be actively compensated using the interference of 'direct' (i.e. order 0) beams. Actually, these beams are unaffected by the RF signals on the AOMs and can be used to from an homodyne interferometer. The LOT interferometer arm lengths can therefore be stabilized using mirror dithering (above 1~kHz, i.e. outside the frequency band of eLISA) and dark fringe stabilization scheme. The homodyne and heterodyne interferometers will share (almost) the same optical paths (they only slightly differ between the AOMs and the end arm mirrors) and the compensation is expected to approach the interferometry requirement of eLISA.

Another effort is currently being made to couple the present command-control system for the LOT, to the LISACode simulation software \cite{Petiteau2008, PetiteauThese2008} also developed in our lab. Once achieved, this work will allow the simulations of realistic propagation delays, taking into account Sagnac effect, variable delays and Doppler shifts. It will also be possible to directly compare the 'numerical' results of TDI (as given by LISACode) to the same algorithm applied on optically or electronically simulated beatnotes. 

Finally, some space have been saved on the optical bench to insert electro-optical modulators. These modulators are planned to simulate the implementation of clock noise transfer and ranging.

\section{Conclusion}

We have reported the progress and first measurements of the eLISA On Table (LOT) experiment), aiming at simulating the optical beatnotes that will be recorded by long-arms space-based interferometers, such as eLISA. The principle of the experiment has been validated, with beatnotes recorded both on optical and electronics interferometers. Presently, time jitter in the generated delays and optical bench phase noise limit the performance of the Time Delay Interferometry (TDI) algorithm to a reduction of about $10^6$ of the noise. Further characterization, hardware and software developments are planned to reduce the residual phase noise level and increase the similarities of the LOT with the expected implementation of eLISA.



%
%

\begin{acknowledgements}
This work has been funded by the French Space Agency (CNES), under grants R-S07/SU-0001-012 and R-S08/SU-0001-012. 
The authors wish to thank G. Heinzel and I. Bykov (Albert Einstein Institut, Hanover, Germany), who provided us the phasemeter and adapted it to our needs.

\end{acknowledgements}

\bibliographystyle{spmpsci}      
\bibliography{LOTDescription_Biblio}   

%
%

\end{document}